\documentclass[%
 aip,
 amsmath,amssymb,
 reprint,%
]{revtex4-1}

\usepackage{graphicx}% Include figure files
\usepackage{dcolumn}% Align table columns on decimal point
\usepackage{bm}% bold math
\usepackage{hyperref}% add hypertext capabilities

\usepackage[utf8]{inputenc}
\usepackage[T1]{fontenc}
\usepackage{mathptmx}
\usepackage{etoolbox}
\usepackage{color}
\makeatletter
\def\@email#1#2{%
 \endgroup
 \patchcmd{\titleblock@produce}
  {\frontmatter@RRAPformat}
  {\frontmatter@RRAPformat{\produce@RRAP{*#1\href{mailto:#2}{#2}}}\frontmatter@RRAPformat}
  {}{}
}%
\makeatother
\begin{document}

\preprint{AIP/123-QED}

\title{Non-zero crossing current-voltage characteristics of interface-type resistive switching devices}

\author{Sahitya Yarragolla}
\email{sahitya.yarragolla@ruhr-uni-bochum.de}
\homepage{https://orcid.org/0000-0002-2973-4943}
\affiliation{Chair of Applied Electrodynamics and Plasma Technology, Ruhr University Bochum, Germany}

\author{Torben Hemke}
\homepage{https://orcid.org/0000-0003-2436-5840}
\affiliation{Chair of Applied Electrodynamics and Plasma Technology, Ruhr University Bochum, Germany}

\author{Jan Trieschmann}
\homepage{https://orcid.org/0000-0001-9136-8019}
\affiliation{Theoretical Electrical Engineering, Faculty of Engineering, Kiel University, Kaiserstraße 2, 24143 Kiel, Germany}
\affiliation{Kiel Nano, Surface and Interface Science KiNSIS, Kiel University, Christian-Albrechts-Platz 4, 24118 Kiel, Germany}

\author{Thomas Mussenbrock}
\email{thomas.mussenbrock@ruhr-uni-bochum.de}
\homepage{https://orcid.org/0000-0001-6445-4990}
\affiliation{Chair of Applied Electrodynamics and Plasma Technology, Ruhr University Bochum, Germany}

\date{\today}

\begin{abstract}
A number of memristive devices, mainly ReRAMs, have been reported to exhibit a unique non-zero crossing hysteresis attributed to the interplay of resistive and not yet fully understood `capacitive', and `inductive' effects. This work exploits a kinetic simulation model based on the stochastic cloud-in-a-cell method to capture these effects. The model, applied to Au/BiFeO$_{3}$/Pt/Ti interface-type devices, incorporates vacancy transport and capacitive contributions. The resulting nonlinear response, characterized by hysteresis, is analyzed in detail, providing an in-depth physical understanding of the virtual effects. Capacitive effects are modeled across different layers, revealing their significant role in shaping the non-zero crossing hysteresis behavior. Results from kinetic simulations demonstrate the impact of frequency-dependent impedance on the non-zero crossing phenomenon. This model provides insights into the effects of various device material properties, such as Schottky barrier height, device area and oxide layer on the non-zero crossing point.
\end{abstract}

\maketitle

%The transition from standard CMOS-based devices to memristive devices in integrated circuits is a revolutionary advancement in electronic devices and computing. For decades, CMOS-based devices have remained the dominant devices in electronics. 
The transition from standard CMOS-based devices to memristive devices is a revolutionary advancement in electronics and computing. For decades, the electronics industry has been dominated by CMOS-based devices. However, memristive devices have emerged as a promising alternative with distinctive attributes\cite{Song2023}. Memristive devices can retain information about the amount of charge that flows through them, as demonstrated in their history-dependent resistance function \cite{Chua2011, Strukov2008, Waser2021}. As charge accumulates, the conductance of the memristor can change and remain altered until a charge reset occurs. One of the main fingerprints of all memristive devices is their pinched current-voltage characteristics (\textit{I}-\textit{V} curves) or hysteresis \cite{Chua2014}. 

%The current-voltage characteristics of memristive devices typically exhibit a pinched hysteresis shape. This arises from the nonlinear dynamics within these devices. Notably, there is a zero crossing point where the current becomes zero at zero input voltage \cite{Chua2014}. However, several `real-world' devices, mainly resistive switching random access (ReRAM) devices, have shown a non-zero crossing hysteresis \cite{Sun2019, Salaoru2014, Du2018, XU2020, Yang2023c}. This indicates the presence of some charge, like in a battery. It has been found from the literature that the occurrence of such non-zero crossing hysteresis could be due to the presence of capacitive and virtual inductive effects. It is important to note that at least the later effects are not due to physical inductances but could be due to different processes that contribute to switching or parasitic effects that have already been observed in practical devices. This means that it would be more appropriate, as mentioned by Qingjiang et al.\cite{Qingjiang2014}, to consider the nonlinear change in resistive switching as a change in the impedance of the memristive devices rather than just memristance. Such a nonlinear interplay of resistive, capacitive, and virtual inductive effects introduces frequency-dependent impedance, giving rise to higher harmonics in the current and altering the phase relationship between voltage and current. 

\begin{figure*}[!t]
    \centering\includegraphics[width=0.73\textwidth]{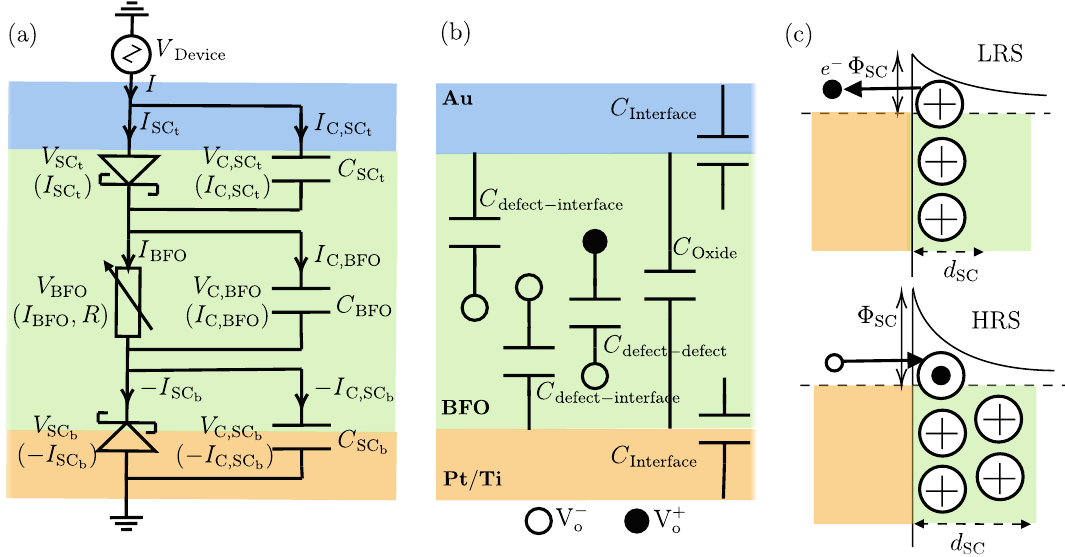}
    \caption{(a) The modified equivalent circuits of BFO device\cite{Yarragolla2022BFO} with parallel capacitors across different layers. (b) Different capacitive components in BFO. (c) Depletion layer width variation at LRS and HRS}
    \label{fig:1}
\end{figure*} 

The \textit{I}-\textit{V} curves of memristive devices typically exhibit a pinched hysteresis shape, arising from their nonlinear dynamics. Notably, there is a zero crossing point where current becomes zero at zero voltage\cite{Chua2014}. However, several `real-world' devices, mainly resistive switching random access (ReRAM) devices, have shown a non-zero crossing hysteresis, indicating presence of some charge, like in a battery\cite{Sun2019, Salaoru2014, Du2018, XU2020, Yang2023c}. Literature suggests that capacitive and virtual inductive effects contribute to this phenomenon. It is important to note that at least the later effects are not due to physical inductances but could be due to different processes that contribute to switching. This means that it would be more appropriate, as mentioned by Qingjiang et al.\cite{Qingjiang2014} to consider the nonlinear change in resistive switching as a change in the impedance rather than just memristance. Such a nonlinear interplay of resistive, capacitive, and virtual inductive effects introduces frequency-dependent impedance, affecting the phase relationship between voltage and current and generating higher harmonics.

%Modeling and simulation techniques provide a powerful tool for understanding the impact of capacitance and virtual inductance on non-zero crossing hysteresis in the \textit{I}-\textit{V} curves of memristive devices. Including ion/vacancy transport and capacitive effects in the modeling of memristive devices is essential. Notably, these models do not include inductive effects artificially; they are included as inertia effects in the particle transport, as explained later. Consequently, we will use the term `inertia effects' rather than `inductive effects'. Several models have been proposed to replicate particle transport-based resistive devices in ReRAMs, which include multidimensional computational models \cite{Dirkmann2016, Dirkmann2018, Funck2021, Aeschlimann2023} and compact models suitable for circuit simulations \cite{Jiang2016, Izquierdo2021, Bengel2020}. Compact models for filamentary devices, including capacitive and/or inductive effects, have been published \cite{Mohamed2015,Qingjiang2014,Berruet2022}. However, such models are currently not available for interface-type ReRAMs. Moreover, existing models do not entirely account for physical and chemical processes contributing to stochastic vacancy transport and virtual effects.
Modeling and simulation techniques provide powerful tools for understanding the impact of capacitive and virtual inductive effects on non-zero crossing hysteresis in I-V curves of memristive devices. Including ion/vacancy transport and capacitive effects in memristive device models is essential, with inductive effects included as inertia effects in particle transport, as explained later. Consequently, we will use the term `inertia effects' rather than `inductive effects'. Several proposed models simulate particle transport-based resistive devices in ReRAMs, offering multidimensional computational \cite{Dirkmann2016, Dirkmann2018, Funck2021, Aeschlimann2023} and compact models suitable for circuit simulations  \cite{Jiang2016, Izquierdo2021, Bengel2020}. Compact models for filamentary devices, including capacitive and/or inductive effects, are published\cite{Mohamed2015,Qingjiang2014,Berruet2022}, such models for interface-type ReRAMs are currently unavailable. Existing models also don't fully account for physical and chemical processes contributing to stochastic vacancy transport and virtual effects.

This paper examines the nonlinear behavior of interface-type memristive devices, utilizing a kinetic simulation model based on the stochastic cloud-in-a-cell (CIC) method for the bismuth ferrite oxide memristive device (BFO) \cite{Yarragolla2022BFO}. Unlike the state-of-the-art compact models, this method more precisely incorporates the stochastic ion or vacancy transport (like the multidimensional computational models), whereas it is fast and accurate like the state-of-the-art compact models. In this paper, the CIC model proposed by Yarragolla et al.\cite{Yarragolla2022DBMD,Yarragolla2022BFO,Yarragolla2023} mainly for interface-type ReRAMs is further modified to incorporate capacitive and inertia effects. 

Typically, the interface-type memristive devices are two-terminal ReRAM devices consisting of a three-layer metal-insulator-metal stack with either a Schottky or tunneling contacts at the metal/oxide interfaces and a solid-state electrolyte sandwiched between these interfaces. The resistive switching in these devices takes place by the drift-diffusion or trapping-de-trapping of charged defects, i.e., positively charged oxygen vacancies in Au/BiFeO$_{3}$/Pt/Ti BFO device \cite{Du2018}. The simulation and modeling of such a mechanism have already been demonstrated for BFO using the CIC method. For a detailed description of the method, refer to the work by Yarragolla et al.\cite{Yarragolla2022DBMD,Yarragolla2022BFO} The pseudo-code of the method and the device-specific parameters used in this work are provided in the supplementary material.\\
           
\paragraph*{Vacancy transport:}

In the CIC-based approach, the solution of particle transport and the electric field are conducted iteratively. For an input voltage bias, the particles are transported based on their drift velocity, calculated using the activation energy, ${U}_{\rm A}$, and electric field, $E$, obtained by solving the Poisson equation. The drift velocity is calculated as follows based on the probability of the particle movement from one lattice site to another \cite{bruce_1994, Meyer2008},
\begin{equation}
 v_{\rm D} = \nu_{0} d \,\, {\rm exp}\left ( -\frac{{U}_{\rm A}}{k_{\rm B}T} \right ) \sinh\left (\frac{\left | z \right |edE}{k_{\rm B}T}  \right ),
 \label{Eq:1}
\end{equation}
where $d$ is the lattice constant, $\nu_{0}$ is the phonon frequency, $k_{B}$ is the Boltzmann constant, $e$ is the elementary charge, $T$ is the temperature, and $z$ is the charge number of the ion. Once the ion or vacancy transport is completed, the electrical parameters, such as the currents and voltages across different layers, are calculated. To incorporate capacitive effects, parallel capacitors are added across different layers in the equivalent circuit models of BFO, as illustrated in Fig.~\ref{fig:1}(a). The modified equivalent circuit yields the following equations by applying Kirchhoff's current and voltage laws:
\begin{equation}
    I_{\rm SC_{t}} + I_{\rm C,SC_{t}} = I_{\rm BFO}+ I_{\rm C,BFO}= -(I_{\rm SC_{b}} + I_{\rm C,SC_{b}}) = I, \hspace{0.2cm} \rm{and}
    \label{Eq:2}
\end{equation}
\begin{equation}
    V_{\rm Device} = V_{\rm SC_{t}} + V_{\rm BFO} - V_{\rm SC_{b}}.
    \label{Eq:3}
\end{equation}
For simplicity, we assume an equal voltage drop across the Schottky contact or oxide, and the capacitors. The resistive current through the Schottky contact is computed as follows \cite{Sze2007},
\begin{equation}
    I_{\rm SC} = A_{d}A^{*} T^{2}{\rm exp}\left \{ \frac{-\Phi_{\rm SC}}{k_{B}T} \right \}\left ( {\rm exp}\left \{ \frac{eV_{\rm SC}}{n_{\rm SC}k_{B}T} \right \} - 1\right ).
    \label{Eq:4}
\end{equation}
\noindent Here $n_{\rm SC}$ is the ideality factor, $\Phi_{\rm SC}$ is the Schottky barrier height, and $A^{*}$ is the effective Richardson constant. Moreover, the current across the oxide region is given by the general Ohm's law,
\begin{equation}
    I_{\rm BFO} = \sigma_{\rm BFO} A_{\rm d}\frac{V_{\rm BFO}}{l_{\rm BFO}},
    \label{Eq:5}
\end{equation}
where $l_{\rm BFO}$ is the length of the active BFO oxide layer and $\sigma$ is its conductivity of BFO. \\

\paragraph*{Capacitive effects:}

%The identification and characterization of capacitive elements in nanoscale ReRAM devices demand a comprehensive investigation of the complex processes within these structures. At this scale, ReRAM devices exhibit complex behaviors that are influenced by many factors, such as metal-insulator interfaces, ion or vacancy transport, and interlayer interactions. The capacitive elements can be distinguished based on the specific functions of various regions inside the device. These regions comprise charge storage at metal-oxide interfaces, capacitance related to defects and interfaces, dielectric layer capacitance, and interlayer capacitance in multi-layered structures. To determine the capacitive components in an interface-type memristive device, Mohamed et al.'s\cite{Mohamed2015} approach for filamentary devices is followed, as illustrated in Fig.~\ref{fig:1}(b). The capacitance in interface-type devices can be composed of four components: the capacitance between the interface and defects (ions or vacancies), the capacitance between positive and negative defects, the oxide capacitance, and the capacitance at the interfaces. In this context, capacitive effects are modeled by applying the effective capacitance value. The methods used to model the capacitance across these interfaces are discussed below.

Identifying and characterizing capacitive elements in nanoscale devices requires an in-depth exploration of the intricate processes within these structures. At this scale, ReRAM devices demonstrate complex behaviors influenced by factors like metal-insulator interfaces, vacancy transport, and interlayer interactions. Capacitive elements within the device can be distinguished based on specific functions in different regions, including charge storage at metal-oxide interfaces, capacitance related to defects and interfaces, dielectric layer capacitance, and interlayer capacitance in multi-layered structures. To determine capacitive components in interface-type devices, Mohamed et al.'s \cite{Mohamed2015} approach for filamentary devices is followed, as shown in Fig. ~\ref{fig:1}(b). This includes capacitance between the interface and defects, capacitance between positive and negative defects, oxide capacitance, and capacitance at the interfaces. Capacitive effects are modeled using effective capacitance values, and the methods for modeling capacitance across these interfaces are discussed below.

\begin{figure}[t]
\centering\includegraphics[width=0.45\textwidth]{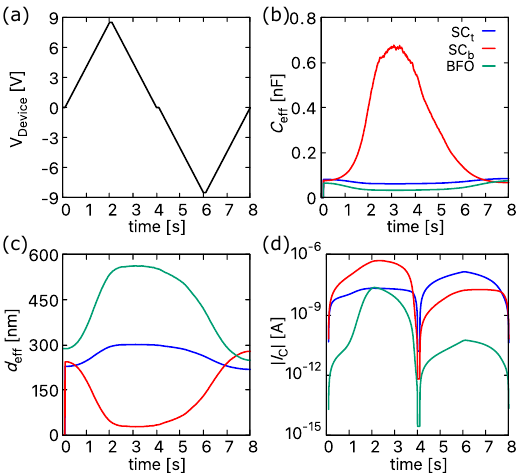}
\caption{The change in (a) input voltage, (b) capacitance, (c) depletion layer width, and (d) capacitive current across the top and bottom Schottky contact.}
\label{fig:2}
\end{figure}

The capacitance of a Schottky contact is affected by the dynamics of the depletion layer, a region near the interface of metal and semiconductors with a lack of charge carriers. Changes in the depletion layer's width occur as charge accumulates or depletes in this region, ultimately affecting the overall capacitance of the Schottky contact \cite{grundmann2015}. This is demonstrated in Fig.~\ref{fig:1}{c}. When a positive voltage is applied to the metal-semiconductor junction, the barrier height decreases, attracting more charge carriers to the interface and causing a narrowing of the depletion layer. Some electrons may flow through the barrier toward the metal electrode, leaving positively charged empty traps near the interface. In contrast, applying a negative bias raises the barrier and prevents electrons from flowing through it. Yet they may be injected into the empty traps. This causes the traps to become neutral and resume functioning, increasing the depletion width \cite{Yan2013}. The change in depletion layer width corresponds to fluctuations in the capacitance of the Schottky contact. The depletion region width can be determined based on the charge accumulation of the effective barrier height at the metal/oxide interfaces as follows:
\begin{equation}
    d_{\rm{SC}} = \sqrt{\frac{2\epsilon_{0}\epsilon_{\textrm{r,BFO}}\left ( \Phi_{\rm{SC}}-qV_{\textrm{SC}}-k_{\textrm{B}}T \right )}{en}},
     \label{Eq:6}
\end{equation}
\begin{equation}
    d_{\rm{SC_{eff}}} = d_{\rm{SC}}(1 + \lambda_{d}\,q(t)),
     \label{Eq:7}
\end{equation}
 where $\epsilon_{0}$ is the vacuum permittivity,$\epsilon_{\textrm{r,BFO}}$ is the relative permittivity of BFO and $n$ is the defect density. Eq. \eqref{Eq:7} defines the rate at which the depletion layer width changes where $\lambda_{\rm d}$ is the fitting parameter between 0 and 1, chosen to match the simulation results with experiments and $q(t) = \frac{{d_{\rm BFO}}(t)-{d_{\rm BFO,initial}}}{\bar{d}_{\rm BFO,initial}}$ is the internal state of the device, where
 \begin{equation}
         d_{\rm{BFO}} = \frac{\sum_{i=1}^{N_{\rm vacancies}}\left ( \bar{x}_{\rm i}-\bar{x}_{\rm SC_{b}}\right )}{N_{\rm vacancies}}. 
         \label{Eq:13}
\end{equation}
Here, $\bar{x}_{i}$ is the position of $i^{\rm th}$ mobile vacancy, $\bar{x}_{\rm SC_{b}}$ is the position of bottom Schottky contact and $N_{\rm vacancies}$ is the number of mobile vacancies.

The capacitance of the Schottky junction can be determined by utilizing the electrostatic capacitance equation applied in a parallel plate capacitor using the permittivity of the oxide layer. Furthermore, the general electrostatic capacitance equation is employed to determine the capacitance across a solid-state electrolyte. Here, the effective capacitance resulting from the capacitance between different charged defects and defects and interface is considered.

\begin{figure}[t]
\centering\includegraphics[width=0.38\textwidth]{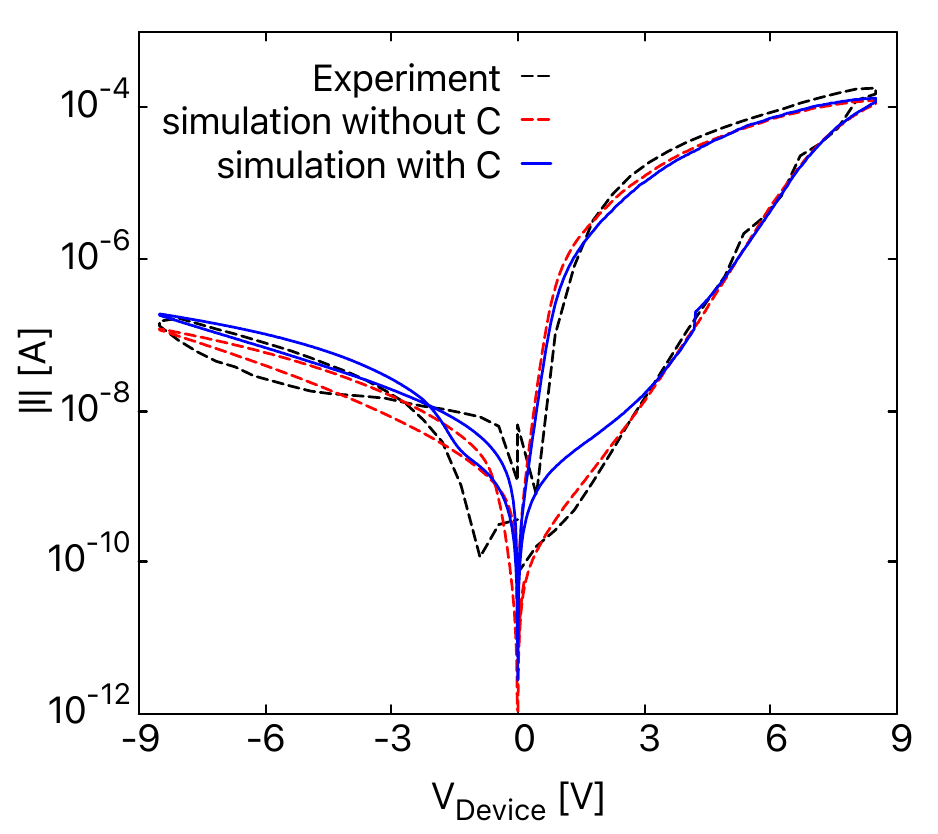}
\caption{The experimental and simulated \textit{I}-\textit{V} curves of BFO memristive devices with and without the capacitive effects (C).}
\label{fig:3}
\end{figure}

The electrostatic capacitance and the current can be calculated using the following formula respectively:
\begin{equation}
    C_{\textrm{SC/BFO}}= \frac{\epsilon_{0}\epsilon_{\textrm{r, BFO}}A_{\textrm{d}} \ r_{\textrm{C}}}{d_{\rm{SC_{eff}}/BFO}} \hspace{0.2cm}  {\rm and}
     \label{Eq:8}
\end{equation}
\begin{equation}
    I_{\rm C,SC/BFO} =  C_{\rm SC/BFO}\frac{\mathrm{d} V_{\textrm{SC/BFO}}}{\mathrm{d} t}
     \label{Eq:9}
\end{equation}
%Capacitance is often adjusted by a roughness factor, which is the deviation of the actual surface area from the ideal geometric surface area. Similarly, a correction factor, $(r_{\rm C})$, is applied to the electrostatic capacitance in ReRAM layers to align simulations with experiments and address nanoscale device non-idealities. The value ranges from 0 to 1, representing capacitance reduction. It considers deviations such as dielectric heterogeneity, interface effects, frequency dependence, oxide thickness variations, quantum effects, and process-related variability. An optimal value is obtained through iterations that minimize disparities between theoretical predictions and experimental results under diverse conditions.
Capacitance is often adjusted by a roughness factor, reflecting deviation of the actual surface area from the ideal geometric area. Similarly, here, a correction factor $(r_{\rm C} \in (0, 1))$ aligns simulations with experiments, addressing nanoscale non-idealities. $r_{\rm C}$ represents capacitance reduction and considers factors like dielectric heterogeneity, interface effects, frequency dependence, oxide thickness variations, quantum effects, and process-related variability. Through iterations, an optimal value of $r_{\rm C}$ minimizes disparities between theoretical predictions and experimental results under diverse conditions.

\begin{figure*}[t]
\centering\includegraphics[width=0.95\textwidth]{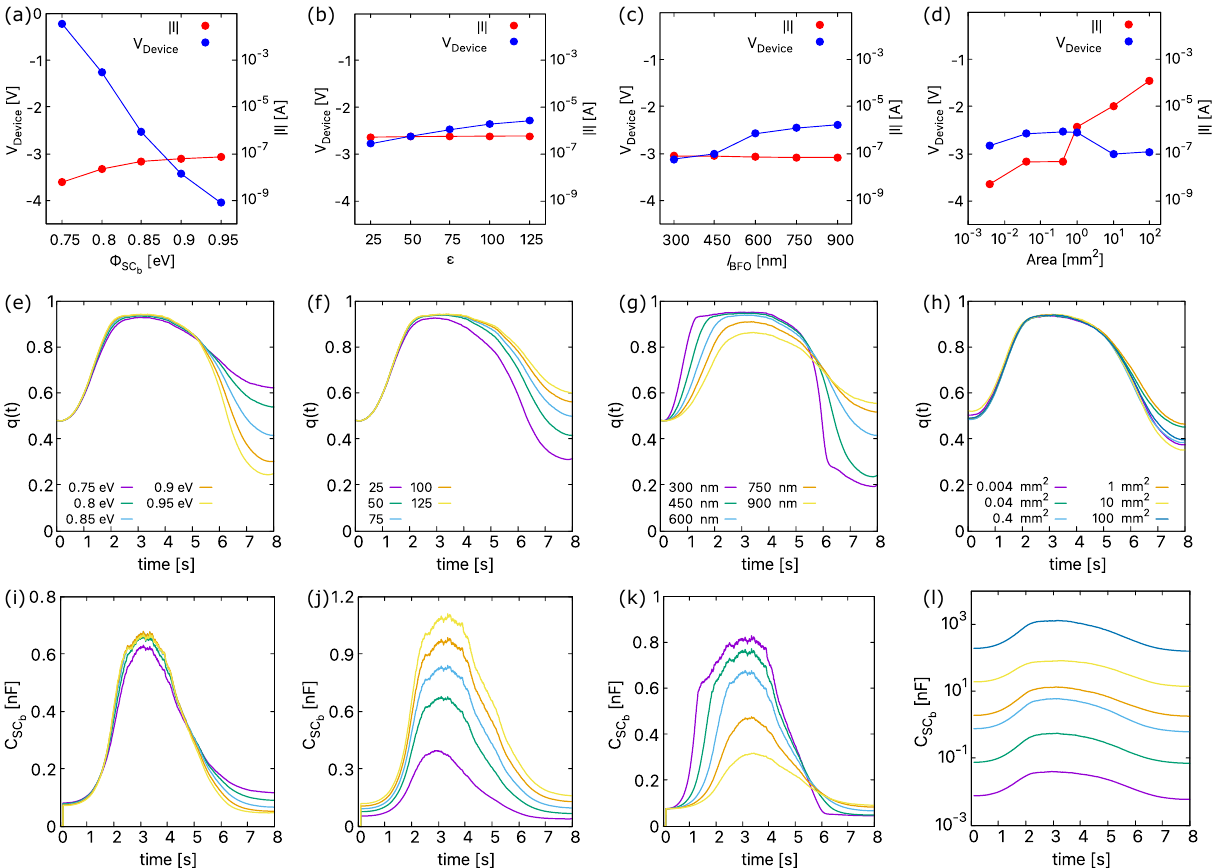}
\caption{(a)-(d) The shift in non-zero crossing point, (e)-(h) the change in $q(t)$ and (i)-(j) the change in bottom Schottky contact capacitance of the device due to the bottom Schottky contact barrier ((a),(e),(i)), BFO permittivity ((b),(f),(j)), BFO layer length ((c),(g),(k)), and the area of the device ((d),(h),(l)). The different colored points in plots (a)-(d) indicate the voltage and current at the non-zero crossing point. }
\label{fig:4}
\end{figure*}

The depletion layer width, its corresponding capacitance, and current across Schottky contact capacitors are displayed in Fig.~\ref{fig:2} for the BFO device. For an input voltage given in Fig.~\ref{fig:2}(a), the top depletion layer width $(d_{\rm SC_{t}})$ near the Au electrode slightly increases while the bottom depletion layer width $(d_{\rm SC_{b}})$ near the Pt electrode decreases as shown in Fig. \ref{fig:2}(c). As oxygen vacancies move towards the Pt electrode, the changes in the top and bottom Schottky barrier heights result in this phenomenon. The capacitance plot (Fig.~\ref{fig:2}(b)) and capacitive current plot (Fig.~\ref{fig:2}(d)) reveal that the capacitance across the bottom Schottky contact has a significant impact on the overall capacitive effects in BFO.

            \begin{figure}[!t]
                \centering      \includegraphics[width=0.45\textwidth]{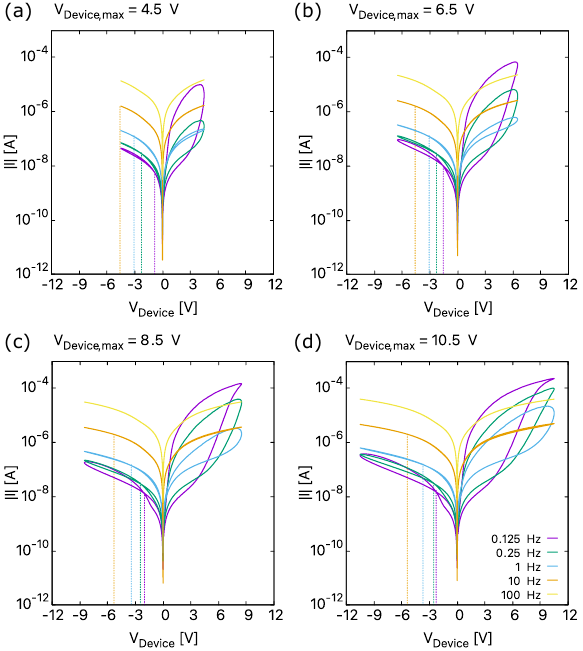}
                \caption{The simulated \textit{I}-\textit{V} curves obtained for a sinusoidal input voltage with various frequencies and amplitudes of (a) 4.5\,V, (b) 6.5\,V, (c) 8.5\,V and (d) 10.5\,V. Dotted lines in each plot indicate the non-zero crossing points. }
                \label{fig:5}
            \end{figure}
\paragraph*{Inertia effects:}
%In solid-state physics and electronic devices, charged particles that participate in resistive switching behaviors imitate electrical circuits. 
The equation that controls the movement of charged particles considering drift velocity and friction, mimics the language of an electrical circuit. This model displays the system as a group of interconnected resistors and inductors, linking particle dynamics with circuit characteristics. This perspective is based on a fundamental approach, i.e., a simplified momentum conservation equation for positive charge carriers,
\begin{equation}
m\frac{\mathrm{d} v_{\mathrm{D}}}{\mathrm{d} t}= e E - m\gamma v_{\mathrm{D}}
\label{Eq:10}
\end{equation}
Here, $v_D$ is the drift velocity, $E$ is the driving electric field, $\gamma$ is the frequency for collisions of charged particles with the atoms of the background lattice. $m$ and $e$ are the particle mass and charge, respectively. After multiplying the momentum equation with $e$ and the particle density $n$ and defining the current density by $j=env_{\mathrm{D}}$, we get a so-called generalized Ohm's law 
\begin{equation}
\frac{\mathrm{d} j}{\mathrm{d} t}= \frac{e^2 n}{m} E - \gamma j
\label{Eq:11}
\end{equation}
In principle, this is nothing but the Drude model of electrical conduction in materials (especially metals). Assuming a homogeneous one-dimensional scenario, we can introduce the current by $I=j A_{\rm d}$ and the voltage by $V=E\,l_{\rm BFO}$. Substituting this, we find
\begin{equation}
V_{\rm BFO}= \frac{m l_{\rm BFO}}{e^2 n A_{\rm d}} \frac{\mathrm{d} I_{\rm BFO}}{\mathrm{d} t} + \frac{m l_{\rm BFO}}{e^2 n A_{\rm d}} \gamma I_{\rm BFO}= L_{\rm BFO} \frac{\mathrm{d} I_{\rm BFO}}{\mathrm{d} t} + R_{\rm BFO} I_{\rm BFO}
\label{Eq:12}
\end{equation}
In this context, $R_{\rm BFO}$ represents ohmic resistance, and $L_{\rm BFO}$ represents an inductance of oxide layer. However, it is important to note that $L_{\rm BFO}$ does not refer to electromagnetic induction. Instead, it serves as a model for inertial effects due to the finite mass of the charged particles involved.

When modeling ion motion in an oxide, it is crucial to allow for the impact of inertia. The momentum equation (generalized Ohm's law) accurately calculates the average velocity of charged particles, comprising ions and electrons, experiencing an electric field.  Consequently, the equation for the voltage drop mirrors the charged particles' movement due to an electric field. It, therefore, includes resistive and inertia effects, and integrating these effects eliminates the need for a separate evaluation when explaining the intricate vacancy movement in an oxide.

By incorporating the above-discussed Eqs. \eqref{Eq:6}-\eqref{Eq:12} into the CIC model for different layers of BFO, the \textit{I}-\textit{V} curves shown in Fig.~\ref{fig:3} are obtained that illustrate the nonlinear behavior of BFO devices resulting from the coexistence of resistive, capacitive, and inertia effects. To obtain the \textit{I}-\textit{V} curves shown in Fig.~\ref{fig:3}, an input voltage (Fig.~\ref{fig:2}(a)) of 8.5\,V was utilized, in conjunction with the parameters used by Yarragolla et al.\cite{Yarragolla2022BFO} in Table~S1. The figure compares the \textit{I}-\textit{V} curves of BFO with and without capacitive effects. %For the red curve, without including capacitive effects, we can still observe a non-zero crossing that occurs at a very low voltage of -0.21\,V. We interpret this as changes in inertial effects attributed to the increased drift velocity or mobility of vacancies during negative bias. 
A non-zero crossing at -0.21\,V is observed for the red curve, even without capacitive effects. Via simulations, we observed that this crossing is random, i.e., we may observe a zero crossing or a non-zero crossing, and it varies between different voltage cycles and devices due to unpredictable stochastic vacancy movement influenced by the electric field and activation energy. Rapid vacancy motion during reset linked to the two-order magnitude increase in vacancy mobility during negative bias, as reported by Du et al. \cite{Du2018}, causes a non-zero crossing in some cycles, highlighting the stochastic nature of vacancy dynamics. As mentioned by Qingjiang et al., inductive effects can also contribute to non-zero crossing \cite{Qingjiang2014}; based on this, we attribute this vacancy motion during RESET to changes in virtual inductivity, which we refer to in this paper as inertial effects.
%As mentioned by Du et al., the mobility of vacancies increases by two orders of magnitude during the negative bias \cite{Du2018}. 
Furthermore, including capacitive effects, the resulting curve (blue) has a similar nonlinear current variation with the input voltage. It reveals similar hysteresis characteristics but with a noticeable non-zero crossing. This curve highlights a precisely similar change in current with the crossing point shifted to -2.9\,V, as observed in the experimental \textit{I}-\textit{V} curve (black). 
            
%For changes in different device parameters, the BFO device \textit{I}-\textit{V} curves show significant shifts in the non-zero crossing point, either vertically (current) or horizontally (voltage).
Analyzing the shifts of non-zero crossing points in BFO device \textit{I}-\textit{V} curves under various parameters provides insights into the interplay of capacitive and inertia effects. These simulation-based findings offer a better understanding of device-switching behavior in response to changes in device parameters, a perspective that may not be easily studied experimentally. An isolated variation of the different capacitive components supports the attribution of the underlying mechanism. As shown in Fig. \ref{fig:4}(a), the capacitance should decrease with higher barrier height, shifting to lower voltages. However, the observed shift in the crossing point to higher $V_{\textrm {Device}}$ is more significant, attributed to changes in inertia effects.  %inductive effects related to vacancy transport, influenced by drift velocity. 
The alterations in the BFO device parameters can affect the electric potential and electric field, affecting vacancies' mobility and drift velocity. This change in drift velocity indirectly affects the inertia effects discussed earlier in Eqs. \eqref{Eq:10}-\eqref{Eq:12}. Consequently, the position of the vacancies and, therefore, $q(t)$ is also altered, which can be considered a way of measuring inertia effects. As illustrated in Fig. \ref{fig:4}(e)-(h) and Fig. \ref{fig:4}(i)-(l), monitoring $q(t)$ and capacitance, respectively, can illuminate these dynamics for a comprehensive understanding. For similar reasons, changes in BFO oxide permittivity shift the crossing point to lower voltages (Fig. \ref{fig:4}(b)) with increased capacitance, but the influence on device operation remains almost constant. Furthermore, variations in BFO layer length as shown in Fig. \ref{fig:4}(c) result in a shift to lower voltages due to the inverse relationship between depletion layer width and capacitance. This shift is primarily due to the interplay of capacitive and inertia effects. Lastly, modifications in the device area lead to a vertical upward shift in currents, as increased capacitance is outweighed by resistive current dominance (Fig. \ref{fig:4}(d)). In summary, the mechanism of non-zero crossing in the overall switching kinetics in ReRAM devices is a complex phenomenon involving resistive, capacitive, and inertia effects (unrelated to any induction). To further support this, we included the corresponding \textit{I}-\textit{V} curves in Fig. S1 and plotted the terms in Eq. \eqref{Eq:12} to demonstrate the inertia effects in Fig. S2 of the supporting material.
 
The frequency-dependent behavior of memristive devices requires a comprehensive exploration due to the notable alterations induced by the introduction of capacitive and inertia effects on their nonlinearity and non-zero crossing hysteresis. Fig.~\ref{fig:5} shows different \textit{I}-\textit{V} curves for a BFO device under sinusoidal input voltages with different amplitudes and frequencies. The plots reveal several observations. Increasing the maximum device voltage $(V_{\rm Device, max})$ at a constant frequency amplifies the hysteresis lobe region, correlating with oxygen vacancy repositioning in the BFO \cite{Izquierdo2021, Yarragolla2022DBMD}. Higher voltages exert stronger forces on vacancies, causing them to drift toward the Pt interface, which alters the $q(t)$ and hence the impedance. This phenomenon is more prevalent at lower frequencies. 

At higher frequencies, the hysteresis loop area decreases, limited by the time for lattice jumps and affecting resistive switching. Beyond 10\,Hz, the \textit{I}-\textit{V} curves exhibit resistor-like behavior in BFO. This can be explained by the dynamic interplay of capacitive and resistive effects at varying frequencies, leading to hysteresis saturation. At increased (decreased) frequencies, the capacitive reactance $(X_{\rm C}=1/2\pi fC)$ rapidly decreases (increases), leading to a swift rise (fall) in capacitive current. Finally, the capacitive effects, identified as the 'battery effect', significantly impact the zero-crossing current. The crossing points deviate from the origin more frequently as the voltage increases \cite{Qingjiang2014}. As capacitance increases with frequency, resulting in reduced capacitive reactance and enhanced charge storage, the system's battery effect intensifies, shifting the crossing point away and leading to a nearly linear current change at 100\,Hz, thereby eliminating the non-zero crossing. Only the parameters for the bottom Schottky contact are plotted in Fig. 4 and Fig. 5 due to their significant influence on the device's switching behavior over the top Schottky contact and oxide layer.

Simulated results confirm capacitive and inertia effects contribute to non-zero crossing hysteresis in the BFO device, particularly noticeable at higher frequencies, adding complexity to its switching characteristics. The research proposes a robust model to comprehend inherent nonlinearity in interface-type ReRAM devices, facilitating frequency response analysis. Incorporating these effects into simulation models for all ReRAM devices is crucial beyond BFO. Understanding derived from such models is vital for advancing research and optimizing ReRAMs for applications in neuromorphic computing and hardware security. The model captures the complex interplay of resistive, capacitive, and inertia effects, providing valuable insights for maximizing ReRAM potential in emerging technologies.
%The simulated results confirm capacitive and inertia (not inductive) effects contributing to the non-zero crossing hysteresis in the BFO device. These effects are more noticeable at higher frequencies and add substantial complexity to the device's switching characteristics. This research proposes a robust model for comprehending the nonlinearity inherent in interface-type ReRAM devices and facilitates frequency response analysis of their performance. Incorporating the aforementioned effects into simulation models for all ReRAM devices is important, extending beyond BFO. The understanding derived from such models is critical to advancing the research and optimization of ReRAMs for applications in neuromorphic computing and hardware security. This model captures the complex interplay of resistive, capacitive, and inertia effects, providing valuable insights for utilizing the full potential of ReRAMs in emerging technologies. 

\section*{Author Contributions}
S. Yarragolla: Methodology, Software, writing – original draft; T. Hemke: Software, Writing – review \& editing; J. Trieschmann: Supervision, Writing – review \& editing; T. Mussenbrock: Methodology, Supervision, Writing – review \& editing.

\section*{Acknowledgements}
Funded by the German Research Foundation (Deutsche Forschungsgemeinschaft, DFG) in the frame of SFB 1461 (Project-ID 434434223) and  - SFB 1461 and Research Grant MU 2332/10-1 (Project-ID 439700144).

\section*{Conflicts of interest}
There are no conflicts to declare.

\section*{Data Availability Statement}
Data available on request from the authors.

\bibliography{aipsamp}% Produces the bibliography via BibTeX.

\begin{thebibliography}{28}%
\makeatletter
\providecommand \@ifxundefined [1]{%
 \@ifx{#1\undefined}
}%
\providecommand \@ifnum [1]{%
 \ifnum #1\expandafter \@firstoftwo
 \else \expandafter \@secondoftwo
 \fi
}%
\providecommand \@ifx [1]{%
 \ifx #1\expandafter \@firstoftwo
 \else \expandafter \@secondoftwo
 \fi
}%
\providecommand \natexlab [1]{#1}%
\providecommand \enquote  [1]{``#1''}%
\providecommand \bibnamefont  [1]{#1}%
\providecommand \bibfnamefont [1]{#1}%
\providecommand \citenamefont [1]{#1}%
\providecommand \href@noop [0]{\@secondoftwo}%
\providecommand \href [0]{\begingroup \@sanitize@url \@href}%
\providecommand \@href[1]{\@@startlink{#1}\@@href}%
\providecommand \@@href[1]{\endgroup#1\@@endlink}%
\providecommand \@sanitize@url [0]{\catcode `\\12\catcode `\$12\catcode `\&12\catcode `\#12\catcode `\^12\catcode `\_12\catcode `\%12\relax}%
\providecommand \@@startlink[1]{}%
\providecommand \@@endlink[0]{}%
\providecommand \url  [0]{\begingroup\@sanitize@url \@url }%
\providecommand \@url [1]{\endgroup\@href {#1}{\urlprefix }}%
\providecommand \urlprefix  [0]{URL }%
\providecommand \Eprint [0]{\href }%
\providecommand \doibase [0]{http://dx.doi.org/}%
\providecommand \selectlanguage [0]{\@gobble}%
\providecommand \bibinfo  [0]{\@secondoftwo}%
\providecommand \bibfield  [0]{\@secondoftwo}%
\providecommand \translation [1]{[#1]}%
\providecommand \BibitemOpen [0]{}%
\providecommand \bibitemStop [0]{}%
\providecommand \bibitemNoStop [0]{.\EOS\space}%
\providecommand \EOS [0]{\spacefactor3000\relax}%
\providecommand \BibitemShut  [1]{\csname bibitem#1\endcsname}%
\let\auto@bib@innerbib\@empty
%</preamble>
\bibitem [{\citenamefont {Song}\ \emph {et~al.}(2023)\citenamefont {Song}, \citenamefont {Kang}, \citenamefont {Zhang}, \citenamefont {Ji}, \citenamefont {Ascoli}, \citenamefont {Messaris}, \citenamefont {Demirkol}, \citenamefont {Dong}, \citenamefont {Aggarwal}, \citenamefont {Wan}, \citenamefont {Hong}, \citenamefont {Cardwell}, \citenamefont {Boybat}, \citenamefont {Seo}, \citenamefont {Lee}, \citenamefont {Lanza}, \citenamefont {Yeon}, \citenamefont {Onen}, \citenamefont {Li}, \citenamefont {Yildiz}, \citenamefont {del Alamo}, \citenamefont {Kim}, \citenamefont {Choi}, \citenamefont {Milano}, \citenamefont {Ricciardi}, \citenamefont {Alff}, \citenamefont {Chai}, \citenamefont {Wang}, \citenamefont {Bhaskaran}, \citenamefont {Hersam}, \citenamefont {Strukov}, \citenamefont {Wong}, \citenamefont {Valov}, \citenamefont {Gao}, \citenamefont {Wu}, \citenamefont {Tetzlaff}, \citenamefont {Sebastian}, \citenamefont {Lu}, \citenamefont {Chua}, \citenamefont {Yang},\ and\ \citenamefont {Kim}}]{Song2023}%
  \BibitemOpen
  \bibfield  {author} {\bibinfo {author} {\bibfnamefont {M.-K.}\ \bibnamefont {Song}}, \bibinfo {author} {\bibfnamefont {J.-H.}\ \bibnamefont {Kang}}, \bibinfo {author} {\bibfnamefont {X.}~\bibnamefont {Zhang}}, \bibinfo {author} {\bibfnamefont {W.}~\bibnamefont {Ji}}, \bibinfo {author} {\bibfnamefont {A.}~\bibnamefont {Ascoli}}, \bibinfo {author} {\bibfnamefont {I.}~\bibnamefont {Messaris}}, \bibinfo {author} {\bibfnamefont {A.~S.}\ \bibnamefont {Demirkol}}, \bibinfo {author} {\bibfnamefont {B.}~\bibnamefont {Dong}}, \bibinfo {author} {\bibfnamefont {S.}~\bibnamefont {Aggarwal}}, \bibinfo {author} {\bibfnamefont {W.}~\bibnamefont {Wan}}, \bibinfo {author} {\bibfnamefont {S.-M.}\ \bibnamefont {Hong}}, \bibinfo {author} {\bibfnamefont {S.~G.}\ \bibnamefont {Cardwell}}, \bibinfo {author} {\bibfnamefont {I.}~\bibnamefont {Boybat}}, \bibinfo {author} {\bibfnamefont {J.-s.}\ \bibnamefont {Seo}}, \bibinfo {author} {\bibfnamefont {J.-S.}\ \bibnamefont {Lee}}, \bibinfo {author} {\bibfnamefont {M.}~\bibnamefont
  {Lanza}}, \bibinfo {author} {\bibfnamefont {H.}~\bibnamefont {Yeon}}, \bibinfo {author} {\bibfnamefont {M.}~\bibnamefont {Onen}}, \bibinfo {author} {\bibfnamefont {J.}~\bibnamefont {Li}}, \bibinfo {author} {\bibfnamefont {B.}~\bibnamefont {Yildiz}}, \bibinfo {author} {\bibfnamefont {J.~A.}\ \bibnamefont {del Alamo}}, \bibinfo {author} {\bibfnamefont {S.}~\bibnamefont {Kim}}, \bibinfo {author} {\bibfnamefont {S.}~\bibnamefont {Choi}}, \bibinfo {author} {\bibfnamefont {G.}~\bibnamefont {Milano}}, \bibinfo {author} {\bibfnamefont {C.}~\bibnamefont {Ricciardi}}, \bibinfo {author} {\bibfnamefont {L.}~\bibnamefont {Alff}}, \bibinfo {author} {\bibfnamefont {Y.}~\bibnamefont {Chai}}, \bibinfo {author} {\bibfnamefont {Z.}~\bibnamefont {Wang}}, \bibinfo {author} {\bibfnamefont {H.}~\bibnamefont {Bhaskaran}}, \bibinfo {author} {\bibfnamefont {M.~C.}\ \bibnamefont {Hersam}}, \bibinfo {author} {\bibfnamefont {D.}~\bibnamefont {Strukov}}, \bibinfo {author} {\bibfnamefont {H.-S.~P.}\ \bibnamefont {Wong}}, \bibinfo
  {author} {\bibfnamefont {I.}~\bibnamefont {Valov}}, \bibinfo {author} {\bibfnamefont {B.}~\bibnamefont {Gao}}, \bibinfo {author} {\bibfnamefont {H.}~\bibnamefont {Wu}}, \bibinfo {author} {\bibfnamefont {R.}~\bibnamefont {Tetzlaff}}, \bibinfo {author} {\bibfnamefont {A.}~\bibnamefont {Sebastian}}, \bibinfo {author} {\bibfnamefont {W.}~\bibnamefont {Lu}}, \bibinfo {author} {\bibfnamefont {L.}~\bibnamefont {Chua}}, \bibinfo {author} {\bibfnamefont {J.~J.}\ \bibnamefont {Yang}}, \ and\ \bibinfo {author} {\bibfnamefont {J.}~\bibnamefont {Kim}},\ }\bibfield  {title} {\enquote {\bibinfo {title} {Recent advances and future prospects for memristive materials, devices, and systems},}\ }\href {\doibase 10.1021/acsnano.3c03505} {\bibfield  {journal} {\bibinfo  {journal} {ACS Nano}\ }\textbf {\bibinfo {volume} {17}},\ \bibinfo {pages} {11994--12039} (\bibinfo {year} {2023})},\ \bibinfo {note} {pMID: 37382380},\ \Eprint {http://arxiv.org/abs/https://doi.org/10.1021/acsnano.3c03505}
  {https://doi.org/10.1021/acsnano.3c03505} \BibitemShut {NoStop}%
\bibitem [{\citenamefont {Chua}(2011)}]{Chua2011}%
  \BibitemOpen
  \bibfield  {author} {\bibinfo {author} {\bibfnamefont {L.}~\bibnamefont {Chua}},\ }\bibfield  {title} {\enquote {\bibinfo {title} {{Resistance switching memories are memristors}},}\ }\href {\doibase 10.1007/s00339-011-6264-9} {\bibfield  {journal} {\bibinfo  {journal} {Applied Physics A}\ }\textbf {\bibinfo {volume} {102}},\ \bibinfo {pages} {765--783} (\bibinfo {year} {2011})}\BibitemShut {NoStop}%
\bibitem [{\citenamefont {Strukov}\ \emph {et~al.}(2008)\citenamefont {Strukov}, \citenamefont {Snider}, \citenamefont {Stewart},\ and\ \citenamefont {Williams}}]{Strukov2008}%
  \BibitemOpen
  \bibfield  {author} {\bibinfo {author} {\bibfnamefont {D.~B.}\ \bibnamefont {Strukov}}, \bibinfo {author} {\bibfnamefont {G.~S.}\ \bibnamefont {Snider}}, \bibinfo {author} {\bibfnamefont {D.~R.}\ \bibnamefont {Stewart}}, \ and\ \bibinfo {author} {\bibfnamefont {R.~S.}\ \bibnamefont {Williams}},\ }\bibfield  {title} {\enquote {\bibinfo {title} {The missing memristor found},}\ }\href {\doibase 10.1038/nature06932} {\bibfield  {journal} {\bibinfo  {journal} {Nature}\ }\textbf {\bibinfo {volume} {453}},\ \bibinfo {pages} {80--83} (\bibinfo {year} {2008})}\BibitemShut {NoStop}%
\bibitem [{\citenamefont {Ielmini}\ and\ \citenamefont {Waser}(2016)}]{Waser2021}%
  \BibitemOpen
  \bibfield  {author} {\bibinfo {author} {\bibfnamefont {D.}~\bibnamefont {Ielmini}}\ and\ \bibinfo {author} {\bibfnamefont {R.}~\bibnamefont {Waser}},\ }\href {\doibase https://doi.org/10.1002/9783527680870} {\emph {\bibinfo {title} {Resistive Switching: From Fundamentals of Nanoionic Redox Processes to Memristive Device Applications}}}\ (\bibinfo  {publisher} {John Wiley \& Sons, Ltd},\ \bibinfo {year} {2016})\BibitemShut {NoStop}%
\bibitem [{\citenamefont {Chua}(2014)}]{Chua2014}%
  \BibitemOpen
  \bibfield  {author} {\bibinfo {author} {\bibfnamefont {L.}~\bibnamefont {Chua}},\ }\bibfield  {title} {\enquote {\bibinfo {title} {If it’s pinched it’s a memristor},}\ }\href {\doibase 10.1088/0268-1242/29/10/104001} {\bibfield  {journal} {\bibinfo  {journal} {Semiconductor Science and Technology}\ }\textbf {\bibinfo {volume} {29}},\ \bibinfo {pages} {104001} (\bibinfo {year} {2014})}\BibitemShut {NoStop}%
\bibitem [{\citenamefont {Yarragolla}\ \emph {et~al.}(2022{\natexlab{a}})\citenamefont {Yarragolla}, \citenamefont {Du}, \citenamefont {Hemke}, \citenamefont {Zhao}, \citenamefont {Chen}, \citenamefont {Polian},\ and\ \citenamefont {Mussenbrock}}]{Yarragolla2022BFO}%
  \BibitemOpen
  \bibfield  {author} {\bibinfo {author} {\bibfnamefont {S.}~\bibnamefont {Yarragolla}}, \bibinfo {author} {\bibfnamefont {N.}~\bibnamefont {Du}}, \bibinfo {author} {\bibfnamefont {T.}~\bibnamefont {Hemke}}, \bibinfo {author} {\bibfnamefont {X.}~\bibnamefont {Zhao}}, \bibinfo {author} {\bibfnamefont {Z.}~\bibnamefont {Chen}}, \bibinfo {author} {\bibfnamefont {I.}~\bibnamefont {Polian}}, \ and\ \bibinfo {author} {\bibfnamefont {T.}~\bibnamefont {Mussenbrock}},\ }\bibfield  {title} {\enquote {\bibinfo {title} {Physics inspired compact modelling of bifeo3 based memristors},}\ }\href {\doibase 10.1038/s41598-022-24439-4} {\bibfield  {journal} {\bibinfo  {journal} {Scientific Reports}\ }\textbf {\bibinfo {volume} {12}},\ \bibinfo {pages} {20490} (\bibinfo {year} {2022}{\natexlab{a}})}\BibitemShut {NoStop}%
\bibitem [{\citenamefont {Sun}\ \emph {et~al.}(2019)\citenamefont {Sun}, \citenamefont {Chen}, \citenamefont {Xiao}, \citenamefont {Zhou}, \citenamefont {Ranjan}, \citenamefont {Hou}, \citenamefont {Zhu}, \citenamefont {Zhao}, \citenamefont {Redfern},\ and\ \citenamefont {Zhou}}]{Sun2019}%
  \BibitemOpen
  \bibfield  {author} {\bibinfo {author} {\bibfnamefont {B.}~\bibnamefont {Sun}}, \bibinfo {author} {\bibfnamefont {Y.}~\bibnamefont {Chen}}, \bibinfo {author} {\bibfnamefont {M.}~\bibnamefont {Xiao}}, \bibinfo {author} {\bibfnamefont {G.}~\bibnamefont {Zhou}}, \bibinfo {author} {\bibfnamefont {S.}~\bibnamefont {Ranjan}}, \bibinfo {author} {\bibfnamefont {W.}~\bibnamefont {Hou}}, \bibinfo {author} {\bibfnamefont {X.}~\bibnamefont {Zhu}}, \bibinfo {author} {\bibfnamefont {Y.}~\bibnamefont {Zhao}}, \bibinfo {author} {\bibfnamefont {S.~A.}\ \bibnamefont {Redfern}}, \ and\ \bibinfo {author} {\bibfnamefont {Y.~N.}\ \bibnamefont {Zhou}},\ }\bibfield  {title} {\enquote {\bibinfo {title} {A unified capacitive-coupled memristive model for the nonpinched current–voltage hysteresis loop},}\ }\href {\doibase 10.1021/acs.nanolett.9b02683} {\bibfield  {journal} {\bibinfo  {journal} {Nano Letters}\ }\textbf {\bibinfo {volume} {19}},\ \bibinfo {pages} {6461--6465} (\bibinfo {year} {2019})},\ \bibinfo {note} {pMID:
  31434487},\ \Eprint {http://arxiv.org/abs/https://doi.org/10.1021/acs.nanolett.9b02683} {https://doi.org/10.1021/acs.nanolett.9b02683} \BibitemShut {NoStop}%
\bibitem [{\citenamefont {Salaoru}\ \emph {et~al.}(2014)\citenamefont {Salaoru}, \citenamefont {Li}, \citenamefont {Khiat},\ and\ \citenamefont {Prodromakis}}]{Salaoru2014}%
  \BibitemOpen
  \bibfield  {author} {\bibinfo {author} {\bibfnamefont {I.}~\bibnamefont {Salaoru}}, \bibinfo {author} {\bibfnamefont {Q.}~\bibnamefont {Li}}, \bibinfo {author} {\bibfnamefont {A.}~\bibnamefont {Khiat}}, \ and\ \bibinfo {author} {\bibfnamefont {T.}~\bibnamefont {Prodromakis}},\ }\bibfield  {title} {\enquote {\bibinfo {title} {Coexistence of memory resistance and memory capacitance in tio2 solid-state devices},}\ }\href {\doibase 10.1186/1556-276X-9-552} {\bibfield  {journal} {\bibinfo  {journal} {Nanoscale Research Letters}\ }\textbf {\bibinfo {volume} {9}},\ \bibinfo {pages} {552} (\bibinfo {year} {2014})}\BibitemShut {NoStop}%
\bibitem [{\citenamefont {Du}\ \emph {et~al.}(2018)\citenamefont {Du}, \citenamefont {Manjunath}, \citenamefont {Li}, \citenamefont {Menzel}, \citenamefont {Linn}, \citenamefont {Waser}, \citenamefont {You}, \citenamefont {B\"urger}, \citenamefont {Skorupa}, \citenamefont {Walczyk}, \citenamefont {Walczyk}, \citenamefont {Schmidt},\ and\ \citenamefont {Schmidt}}]{Du2018}%
  \BibitemOpen
  \bibfield  {author} {\bibinfo {author} {\bibfnamefont {N.}~\bibnamefont {Du}}, \bibinfo {author} {\bibfnamefont {N.}~\bibnamefont {Manjunath}}, \bibinfo {author} {\bibfnamefont {Y.}~\bibnamefont {Li}}, \bibinfo {author} {\bibfnamefont {S.}~\bibnamefont {Menzel}}, \bibinfo {author} {\bibfnamefont {E.}~\bibnamefont {Linn}}, \bibinfo {author} {\bibfnamefont {R.}~\bibnamefont {Waser}}, \bibinfo {author} {\bibfnamefont {T.}~\bibnamefont {You}}, \bibinfo {author} {\bibfnamefont {D.}~\bibnamefont {B\"urger}}, \bibinfo {author} {\bibfnamefont {I.}~\bibnamefont {Skorupa}}, \bibinfo {author} {\bibfnamefont {D.}~\bibnamefont {Walczyk}}, \bibinfo {author} {\bibfnamefont {C.}~\bibnamefont {Walczyk}}, \bibinfo {author} {\bibfnamefont {O.~G.}\ \bibnamefont {Schmidt}}, \ and\ \bibinfo {author} {\bibfnamefont {H.}~\bibnamefont {Schmidt}},\ }\bibfield  {title} {\enquote {\bibinfo {title} {Field-driven hopping transport of oxygen vacancies in memristive oxide switches with interface-mediated resistive switching},}\ }\href
  {\doibase 10.1103/PhysRevApplied.10.054025} {\bibfield  {journal} {\bibinfo  {journal} {Phys. Rev. Applied}\ }\textbf {\bibinfo {volume} {10}},\ \bibinfo {pages} {054025} (\bibinfo {year} {2018})}\BibitemShut {NoStop}%
\bibitem [{\citenamefont {Xu}\ \emph {et~al.}(2020)\citenamefont {Xu}, \citenamefont {Tan}, \citenamefont {Sun}, \citenamefont {Lei}, \citenamefont {Zhao}, \citenamefont {Li}, \citenamefont {Zheng}, \citenamefont {Zhu}, \citenamefont {Zhang},\ and\ \citenamefont {Zhao}}]{XU2020}%
  \BibitemOpen
  \bibfield  {author} {\bibinfo {author} {\bibfnamefont {Y.}~\bibnamefont {Xu}}, \bibinfo {author} {\bibfnamefont {L.}~\bibnamefont {Tan}}, \bibinfo {author} {\bibfnamefont {B.}~\bibnamefont {Sun}}, \bibinfo {author} {\bibfnamefont {M.}~\bibnamefont {Lei}}, \bibinfo {author} {\bibfnamefont {Y.}~\bibnamefont {Zhao}}, \bibinfo {author} {\bibfnamefont {T.}~\bibnamefont {Li}}, \bibinfo {author} {\bibfnamefont {L.}~\bibnamefont {Zheng}}, \bibinfo {author} {\bibfnamefont {S.}~\bibnamefont {Zhu}}, \bibinfo {author} {\bibfnamefont {Y.}~\bibnamefont {Zhang}}, \ and\ \bibinfo {author} {\bibfnamefont {Y.}~\bibnamefont {Zhao}},\ }\bibfield  {title} {\enquote {\bibinfo {title} {Memristive effect with non-zero-crossing current-voltage hysteresis behavior based on ag doped lophatherum gracile brongn},}\ }\href {\doibase https://doi.org/10.1016/j.cap.2020.02.002} {\bibfield  {journal} {\bibinfo  {journal} {Current Applied Physics}\ }\textbf {\bibinfo {volume} {20}},\ \bibinfo {pages} {545--549} (\bibinfo {year}
  {2020})}\BibitemShut {NoStop}%
\bibitem [{\citenamefont {Yang}\ \emph {et~al.}(2023)\citenamefont {Yang}, \citenamefont {Sun}, \citenamefont {Zhou}, \citenamefont {Zhao}, \citenamefont {Zhu}, \citenamefont {Ke}, \citenamefont {Zhao},\ and\ \citenamefont {Wang}}]{Yang2023c}%
  \BibitemOpen
  \bibfield  {author} {\bibinfo {author} {\bibfnamefont {C.}~\bibnamefont {Yang}}, \bibinfo {author} {\bibfnamefont {B.}~\bibnamefont {Sun}}, \bibinfo {author} {\bibfnamefont {G.}~\bibnamefont {Zhou}}, \bibinfo {author} {\bibfnamefont {H.}~\bibnamefont {Zhao}}, \bibinfo {author} {\bibfnamefont {S.}~\bibnamefont {Zhu}}, \bibinfo {author} {\bibfnamefont {C.}~\bibnamefont {Ke}}, \bibinfo {author} {\bibfnamefont {Y.}~\bibnamefont {Zhao}}, \ and\ \bibinfo {author} {\bibfnamefont {H.}~\bibnamefont {Wang}},\ }\bibfield  {title} {\enquote {\bibinfo {title} {Evolution between volatile and nonvolatile resistive switching behaviors in ag/tiox/ceoy/f-doped sno2 nanostructure-based memristor devices for information processing applications},}\ }\href {\doibase 10.1021/acsanm.3c01282} {\bibfield  {journal} {\bibinfo  {journal} {ACS Applied Nano Materials}\ }\textbf {\bibinfo {volume} {6}},\ \bibinfo {pages} {8857--8867} (\bibinfo {year} {2023})},\ \Eprint {http://arxiv.org/abs/https://doi.org/10.1021/acsanm.3c01282}
  {https://doi.org/10.1021/acsanm.3c01282} \BibitemShut {NoStop}%
\bibitem [{\citenamefont {Qingjiang}\ \emph {et~al.}(2014)\citenamefont {Qingjiang}, \citenamefont {Khiat}, \citenamefont {Salaoru}, \citenamefont {Papavassiliou}, \citenamefont {Hui},\ and\ \citenamefont {Prodromakis}}]{Qingjiang2014}%
  \BibitemOpen
  \bibfield  {author} {\bibinfo {author} {\bibfnamefont {L.}~\bibnamefont {Qingjiang}}, \bibinfo {author} {\bibfnamefont {A.}~\bibnamefont {Khiat}}, \bibinfo {author} {\bibfnamefont {I.}~\bibnamefont {Salaoru}}, \bibinfo {author} {\bibfnamefont {C.}~\bibnamefont {Papavassiliou}}, \bibinfo {author} {\bibfnamefont {X.}~\bibnamefont {Hui}}, \ and\ \bibinfo {author} {\bibfnamefont {T.}~\bibnamefont {Prodromakis}},\ }\bibfield  {title} {\enquote {\bibinfo {title} {Memory impedance in tio2 based metal-insulator-metal devices},}\ }\href {\doibase 10.1038/srep04522} {\bibfield  {journal} {\bibinfo  {journal} {Scientific Reports}\ }\textbf {\bibinfo {volume} {4}},\ \bibinfo {pages} {4522} (\bibinfo {year} {2014})}\BibitemShut {NoStop}%
\bibitem [{\citenamefont {Dirkmann}\ \emph {et~al.}(2016)\citenamefont {Dirkmann}, \citenamefont {Hansen}, \citenamefont {Ziegler}, \citenamefont {Kohlstedt},\ and\ \citenamefont {Mussenbrock}}]{Dirkmann2016}%
  \BibitemOpen
  \bibfield  {author} {\bibinfo {author} {\bibfnamefont {S.}~\bibnamefont {Dirkmann}}, \bibinfo {author} {\bibfnamefont {M.}~\bibnamefont {Hansen}}, \bibinfo {author} {\bibfnamefont {M.}~\bibnamefont {Ziegler}}, \bibinfo {author} {\bibfnamefont {H.}~\bibnamefont {Kohlstedt}}, \ and\ \bibinfo {author} {\bibfnamefont {T.}~\bibnamefont {Mussenbrock}},\ }\bibfield  {title} {\enquote {\bibinfo {title} {{The role of ion transport phenomena in memristive double barrier devices}},}\ }\href {\doibase 10.1038/srep35686} {\bibfield  {journal} {\bibinfo  {journal} {Scientific Reports}\ }\textbf {\bibinfo {volume} {6}},\ \bibinfo {pages} {35686} (\bibinfo {year} {2016})}\BibitemShut {NoStop}%
\bibitem [{\citenamefont {Dirkmann}\ \emph {et~al.}(2018)\citenamefont {Dirkmann}, \citenamefont {Kaiser}, \citenamefont {Wenger},\ and\ \citenamefont {Mussenbrock}}]{Dirkmann2018}%
  \BibitemOpen
  \bibfield  {author} {\bibinfo {author} {\bibfnamefont {S.}~\bibnamefont {Dirkmann}}, \bibinfo {author} {\bibfnamefont {J.}~\bibnamefont {Kaiser}}, \bibinfo {author} {\bibfnamefont {C.}~\bibnamefont {Wenger}}, \ and\ \bibinfo {author} {\bibfnamefont {T.}~\bibnamefont {Mussenbrock}},\ }\bibfield  {title} {\enquote {\bibinfo {title} {{Filament Growth and Resistive Switching in Hafnium Oxide Memristive Devices}},}\ }\href {\doibase 10.1021/acsami.7b19836} {\bibfield  {journal} {\bibinfo  {journal} {ACS Applied Materials and Interfaces}\ }\textbf {\bibinfo {volume} {10}},\ \bibinfo {pages} {14857--14868} (\bibinfo {year} {2018})}\BibitemShut {NoStop}%
\bibitem [{\citenamefont {Funck}\ and\ \citenamefont {Menzel}(2021)}]{Funck2021}%
  \BibitemOpen
  \bibfield  {author} {\bibinfo {author} {\bibfnamefont {C.}~\bibnamefont {Funck}}\ and\ \bibinfo {author} {\bibfnamefont {S.}~\bibnamefont {Menzel}},\ }\bibfield  {title} {\enquote {\bibinfo {title} {Comprehensive model of electron conduction in oxide-based memristive devices},}\ }\href {\doibase 10.1021/acsaelm.1c00398} {\bibfield  {journal} {\bibinfo  {journal} {ACS Applied Electronic Materials}\ }\textbf {\bibinfo {volume} {3}},\ \bibinfo {pages} {3674--3692} (\bibinfo {year} {2021})},\ \Eprint {http://arxiv.org/abs/https://doi.org/10.1021/acsaelm.1c00398} {https://doi.org/10.1021/acsaelm.1c00398} \BibitemShut {NoStop}%
\bibitem [{\citenamefont {Aeschlimann}\ \emph {et~al.}(2023)\citenamefont {Aeschlimann}, \citenamefont {Ducry}, \citenamefont {Weilenmann}, \citenamefont {Leuthold}, \citenamefont {Emboras},\ and\ \citenamefont {Luisier}}]{Aeschlimann2023}%
  \BibitemOpen
  \bibfield  {author} {\bibinfo {author} {\bibfnamefont {J.}~\bibnamefont {Aeschlimann}}, \bibinfo {author} {\bibfnamefont {F.}~\bibnamefont {Ducry}}, \bibinfo {author} {\bibfnamefont {C.}~\bibnamefont {Weilenmann}}, \bibinfo {author} {\bibfnamefont {J.}~\bibnamefont {Leuthold}}, \bibinfo {author} {\bibfnamefont {A.}~\bibnamefont {Emboras}}, \ and\ \bibinfo {author} {\bibfnamefont {M.}~\bibnamefont {Luisier}},\ }\bibfield  {title} {\enquote {\bibinfo {title} {Multiscale modeling of metal-oxide-metal conductive bridging random-access memory cells: From ab initio to finite-element calculations},}\ }\href {\doibase 10.1103/PhysRevApplied.19.024058} {\bibfield  {journal} {\bibinfo  {journal} {Phys. Rev. Appl.}\ }\textbf {\bibinfo {volume} {19}},\ \bibinfo {pages} {024058} (\bibinfo {year} {2023})}\BibitemShut {NoStop}%
\bibitem [{\citenamefont {Jiang}\ \emph {et~al.}(2016)\citenamefont {Jiang}, \citenamefont {Wu}, \citenamefont {Yu}, \citenamefont {Yang}, \citenamefont {Song}, \citenamefont {Karim},\ and\ \citenamefont {Wong}}]{Jiang2016}%
  \BibitemOpen
  \bibfield  {author} {\bibinfo {author} {\bibfnamefont {Z.}~\bibnamefont {Jiang}}, \bibinfo {author} {\bibfnamefont {Y.}~\bibnamefont {Wu}}, \bibinfo {author} {\bibfnamefont {S.}~\bibnamefont {Yu}}, \bibinfo {author} {\bibfnamefont {L.}~\bibnamefont {Yang}}, \bibinfo {author} {\bibfnamefont {K.}~\bibnamefont {Song}}, \bibinfo {author} {\bibfnamefont {Z.}~\bibnamefont {Karim}}, \ and\ \bibinfo {author} {\bibfnamefont {H.-S.~P.}\ \bibnamefont {Wong}},\ }\bibfield  {title} {\enquote {\bibinfo {title} {A compact model for metal–oxide resistive random access memory with experiment verification},}\ }\href {\doibase 10.1109/TED.2016.2545412} {\bibfield  {journal} {\bibinfo  {journal} {IEEE Transactions on Electron Devices}\ }\textbf {\bibinfo {volume} {63}},\ \bibinfo {pages} {1884--1892} (\bibinfo {year} {2016})}\BibitemShut {NoStop}%
\bibitem [{\citenamefont {Maestro-Izquierdo}\ \emph {et~al.}(2021)\citenamefont {Maestro-Izquierdo}, \citenamefont {Gonzalez}, \citenamefont {Campabadal}, \citenamefont {Suñé},\ and\ \citenamefont {Miranda}}]{Izquierdo2021}%
  \BibitemOpen
  \bibfield  {author} {\bibinfo {author} {\bibfnamefont {M.}~\bibnamefont {Maestro-Izquierdo}}, \bibinfo {author} {\bibfnamefont {M.~B.}\ \bibnamefont {Gonzalez}}, \bibinfo {author} {\bibfnamefont {F.}~\bibnamefont {Campabadal}}, \bibinfo {author} {\bibfnamefont {J.}~\bibnamefont {Suñé}}, \ and\ \bibinfo {author} {\bibfnamefont {E.}~\bibnamefont {Miranda}},\ }\bibfield  {title} {\enquote {\bibinfo {title} {A new perspective towards the understanding of the frequency-dependent behavior of memristive devices},}\ }\href {\doibase 10.1109/LED.2021.3063239} {\bibfield  {journal} {\bibinfo  {journal} {IEEE Electron Device Letters}\ }\textbf {\bibinfo {volume} {42}},\ \bibinfo {pages} {565--568} (\bibinfo {year} {2021})}\BibitemShut {NoStop}%
\bibitem [{\citenamefont {Bengel}\ \emph {et~al.}(2020)\citenamefont {Bengel}, \citenamefont {Siemon}, \citenamefont {Cüppers}, \citenamefont {Hoffmann-Eifert}, \citenamefont {Hardtdegen}, \citenamefont {von Witzleben}, \citenamefont {Hellmich}, \citenamefont {Waser},\ and\ \citenamefont {Menzel}}]{Bengel2020}%
  \BibitemOpen
  \bibfield  {author} {\bibinfo {author} {\bibfnamefont {C.}~\bibnamefont {Bengel}}, \bibinfo {author} {\bibfnamefont {A.}~\bibnamefont {Siemon}}, \bibinfo {author} {\bibfnamefont {F.}~\bibnamefont {Cüppers}}, \bibinfo {author} {\bibfnamefont {S.}~\bibnamefont {Hoffmann-Eifert}}, \bibinfo {author} {\bibfnamefont {A.}~\bibnamefont {Hardtdegen}}, \bibinfo {author} {\bibfnamefont {M.}~\bibnamefont {von Witzleben}}, \bibinfo {author} {\bibfnamefont {L.}~\bibnamefont {Hellmich}}, \bibinfo {author} {\bibfnamefont {R.}~\bibnamefont {Waser}}, \ and\ \bibinfo {author} {\bibfnamefont {S.}~\bibnamefont {Menzel}},\ }\bibfield  {title} {\enquote {\bibinfo {title} {Variability-aware modeling of filamentary oxide-based bipolar resistive switching cells using spice level compact models},}\ }\href {\doibase 10.1109/TCSI.2020.3018502} {\bibfield  {journal} {\bibinfo  {journal} {IEEE Transactions on Circuits and Systems I: Regular Papers}\ }\textbf {\bibinfo {volume} {67}},\ \bibinfo {pages} {4618--4630} (\bibinfo {year}
  {2020})}\BibitemShut {NoStop}%
\bibitem [{\citenamefont {Mohamed}, \citenamefont {Kim},\ and\ \citenamefont {Cho}(2015)}]{Mohamed2015}%
  \BibitemOpen
  \bibfield  {author} {\bibinfo {author} {\bibfnamefont {M.~G.~A.}\ \bibnamefont {Mohamed}}, \bibinfo {author} {\bibfnamefont {H.}~\bibnamefont {Kim}}, \ and\ \bibinfo {author} {\bibfnamefont {T.-W.}\ \bibnamefont {Cho}},\ }\bibfield  {title} {\enquote {\bibinfo {title} {Modeling of memristive and memcapacitive behaviors in metal-oxide junctions},}\ }\href {\doibase https://doi.org/10.1155/2015/910126} {\bibfield  {journal} {\bibinfo  {journal} {ScientificWorldJournal}\ }\textbf {\bibinfo {volume} {2015}},\ \bibinfo {pages} {910126} (\bibinfo {year} {2015})}\BibitemShut {NoStop}%
\bibitem [{\citenamefont {Berruet}\ \emph {et~al.}(2022)\citenamefont {Berruet}, \citenamefont {Pérez-Martínez}, \citenamefont {Romero}, \citenamefont {Gonzales}, \citenamefont {Al-Mayouf}, \citenamefont {Guerrero},\ and\ \citenamefont {Bisquert}}]{Berruet2022}%
  \BibitemOpen
  \bibfield  {author} {\bibinfo {author} {\bibfnamefont {M.}~\bibnamefont {Berruet}}, \bibinfo {author} {\bibfnamefont {J.~C.}\ \bibnamefont {Pérez-Martínez}}, \bibinfo {author} {\bibfnamefont {B.}~\bibnamefont {Romero}}, \bibinfo {author} {\bibfnamefont {C.}~\bibnamefont {Gonzales}}, \bibinfo {author} {\bibfnamefont {A.~M.}\ \bibnamefont {Al-Mayouf}}, \bibinfo {author} {\bibfnamefont {A.}~\bibnamefont {Guerrero}}, \ and\ \bibinfo {author} {\bibfnamefont {J.}~\bibnamefont {Bisquert}},\ }\bibfield  {title} {\enquote {\bibinfo {title} {Physical model for the current–voltage hysteresis and impedance of halide perovskite memristors},}\ }\href {\doibase 10.1021/acsenergylett.2c00121} {\bibfield  {journal} {\bibinfo  {journal} {ACS Energy Letters}\ }\textbf {\bibinfo {volume} {7}},\ \bibinfo {pages} {1214--1222} (\bibinfo {year} {2022})},\ \Eprint {http://arxiv.org/abs/https://doi.org/10.1021/acsenergylett.2c00121} {https://doi.org/10.1021/acsenergylett.2c00121} \BibitemShut {NoStop}%
\bibitem [{\citenamefont {Yarragolla}\ \emph {et~al.}(2022{\natexlab{b}})\citenamefont {Yarragolla}, \citenamefont {Hemke}, \citenamefont {Trieschmann}, \citenamefont {Zahari}, \citenamefont {Kohlstedt},\ and\ \citenamefont {Mussenbrock}}]{Yarragolla2022DBMD}%
  \BibitemOpen
  \bibfield  {author} {\bibinfo {author} {\bibfnamefont {S.}~\bibnamefont {Yarragolla}}, \bibinfo {author} {\bibfnamefont {T.}~\bibnamefont {Hemke}}, \bibinfo {author} {\bibfnamefont {J.}~\bibnamefont {Trieschmann}}, \bibinfo {author} {\bibfnamefont {F.}~\bibnamefont {Zahari}}, \bibinfo {author} {\bibfnamefont {H.}~\bibnamefont {Kohlstedt}}, \ and\ \bibinfo {author} {\bibfnamefont {T.}~\bibnamefont {Mussenbrock}},\ }\bibfield  {title} {\enquote {\bibinfo {title} {Stochastic behavior of an interface-based memristive device},}\ }\href {\doibase 10.1063/5.0084085} {\bibfield  {journal} {\bibinfo  {journal} {Journal of Applied Physics}\ }\textbf {\bibinfo {volume} {131}},\ \bibinfo {pages} {134304} (\bibinfo {year} {2022}{\natexlab{b}})}\BibitemShut {NoStop}%
\bibitem [{\citenamefont {Yarragolla}, \citenamefont {Hemke},\ and\ \citenamefont {Mussenbrock}(2023)}]{Yarragolla2023}%
  \BibitemOpen
  \bibfield  {author} {\bibinfo {author} {\bibfnamefont {S.}~\bibnamefont {Yarragolla}}, \bibinfo {author} {\bibfnamefont {T.}~\bibnamefont {Hemke}}, \ and\ \bibinfo {author} {\bibfnamefont {T.}~\bibnamefont {Mussenbrock}},\ }\bibfield  {title} {\enquote {\bibinfo {title} {A generic compact and stochastic model for non-filamentary analog resistive switching devices},}\ }in\ \href {\doibase 10.1109/MOCAST57943.2023.10176574} {\emph {\bibinfo {booktitle} {2023 12th International Conference on Modern Circuits and Systems Technologies (MOCAST)}}}\ (\bibinfo {year} {2023})\ pp.\ \bibinfo {pages} {1--4}\BibitemShut {NoStop}%
\bibitem [{\citenamefont {Bruce}(1994)}]{bruce_1994}%
  \BibitemOpen
  \bibfield  {author} {\bibinfo {author} {\bibfnamefont {P.~G.}\ \bibnamefont {Bruce}},\ }\href {\doibase 10.1017/CBO9780511524790} {\emph {\bibinfo {title} {Solid State Electrochemistry}}},\ Chemistry of Solid State Materials\ (\bibinfo  {publisher} {Cambridge University Press},\ \bibinfo {year} {1994})\ Chap.~\bibinfo {chapter} {3}\BibitemShut {NoStop}%
\bibitem [{\citenamefont {Meyer}\ \emph {et~al.}(2008)\citenamefont {Meyer}, \citenamefont {Schloss}, \citenamefont {Brewer}, \citenamefont {Lambertson}, \citenamefont {Kinney}, \citenamefont {Sanchez},\ and\ \citenamefont {Rinerson}}]{Meyer2008}%
  \BibitemOpen
  \bibfield  {author} {\bibinfo {author} {\bibfnamefont {R.}~\bibnamefont {Meyer}}, \bibinfo {author} {\bibfnamefont {L.}~\bibnamefont {Schloss}}, \bibinfo {author} {\bibfnamefont {J.}~\bibnamefont {Brewer}}, \bibinfo {author} {\bibfnamefont {R.}~\bibnamefont {Lambertson}}, \bibinfo {author} {\bibfnamefont {W.}~\bibnamefont {Kinney}}, \bibinfo {author} {\bibfnamefont {J.}~\bibnamefont {Sanchez}}, \ and\ \bibinfo {author} {\bibfnamefont {D.}~\bibnamefont {Rinerson}},\ }\bibfield  {title} {\enquote {\bibinfo {title} {{Oxide dual-layer memory element for scalable non-volatile cross-point memory technology}},}\ }in\ \href {\doibase 10.1109/NVMT.2008.4731194} {\emph {\bibinfo {booktitle} {Proceedings - 9th Annual Non-Volatile Memory Technology Symposium, NVMTS}}}\ (\bibinfo {year} {2008})\BibitemShut {NoStop}%
\bibitem [{\citenamefont {Sze}\ and\ \citenamefont {Ng}(2007)}]{Sze2007}%
  \BibitemOpen
  \bibfield  {author} {\bibinfo {author} {\bibfnamefont {S.~M.}\ \bibnamefont {Sze}}\ and\ \bibinfo {author} {\bibfnamefont {K.~K.}\ \bibnamefont {Ng}},\ }\href {\doibase https://doi.org/10.1002/0470068329} {\emph {\bibinfo {title} {Physics of Semiconductor Devices}}}\ (\bibinfo  {publisher} {John Wiley \& Sons Ltd,},\ \bibinfo {year} {2007})\BibitemShut {NoStop}%
\bibitem [{\citenamefont {Grundmann}(2015)}]{grundmann2015}%
  \BibitemOpen
  \bibfield  {author} {\bibinfo {author} {\bibfnamefont {M.}~\bibnamefont {Grundmann}},\ }\href {https://books.google.de/books?id=VEdECwAAQBAJ} {\emph {\bibinfo {title} {The Physics of Semiconductors: An Introduction Including Nanophysics and Applications}}},\ Graduate Texts in Physics\ (\bibinfo  {publisher} {Springer International Publishing},\ \bibinfo {year} {2015})\BibitemShut {NoStop}%
\bibitem [{\citenamefont {Yan}\ and\ \citenamefont {Liu}(2013)}]{Yan2013}%
  \BibitemOpen
  \bibfield  {author} {\bibinfo {author} {\bibfnamefont {Z.~B.}\ \bibnamefont {Yan}}\ and\ \bibinfo {author} {\bibfnamefont {J.-M.}\ \bibnamefont {Liu}},\ }\bibfield  {title} {\enquote {\bibinfo {title} {Coexistence of high performance resistance and capacitance memory based on multilayered metal-oxide structures},}\ }\href {\doibase 10.1038/srep02482} {\bibfield  {journal} {\bibinfo  {journal} {Scientific Reports}\ }\textbf {\bibinfo {volume} {3}},\ \bibinfo {pages} {2482} (\bibinfo {year} {2013})}\BibitemShut {NoStop}%
\end{thebibliography}%

\end{document}

\section*{Supplementary material}

\vspace{0.5cm}
\noindent
{\large \textbf{Non-zero crossing current-voltage characteristics of interface-type memristive devices}}

\vspace{1cm}
\subsection*{Device parameters}

The BFO device parameters used to obtain plots in Fig. 2-5.

\begin{table}[!ht]
\centering
\begin{tabular}{|l|c|c|}
\hline
\textbf{Physical quantity} & \textbf{Symbol} & \textbf{Value} \\
\hline
Temperature (K) & $T$ & 298 \\
\hline
Phonon frequency (Hz) & $\nu_{0}$ & $1 \times 10^{12}$\\
\hline
Lattice constant (m) & $d$ & $0.56 \times 10^{-9}$\\
\hline
Device area ($\rm mm^{2}$) & $A_{\rm d}$ & $0.04$\\
\hline
Relative permittivity of $\rm BiFeO_{3}$ & $\varepsilon_{r} $ & 52.0\\
\hline
Conductivity of $\rm BiFeO_{3}$ ($\Omega {\rm m}$) & $\sigma$ & $7.0 \times
 10^{-4}$\\
 \hline
 Length of BFO (m)& $l_{\rm BFO}$ & $600 \times 10^{-9}$\\
 \hline
Defect density (${\rm cm^{-3}}$) & $\rho$ & $8\times 10^{16}  $\\
\hline
Top Schottky barrier height (eV)& $\Phi^{\rm t}_{\rm 0}$ & $0.75$\\
\hline
Top Schottky barrier ideality factor & $n_{0}^{\rm t}$ & 4.2\\
\hline
Bottom Schottky barrier height (eV) & $\Phi^{\rm b}_{\rm 0}$ & $0.85$\\
\hline
Bottom Schottky barrier ideality factor & $n_{0}^{\rm b}$ & 4.5\\
\hline
\end{tabular}
%\caption{\label{tab:Sim1}Details of the simulation parameters~\cite{Yarragolla2022}}
\caption*{Table S1: Details of the simulation parameters~\cite{Yarragolla2022}}
\label{tab:S1}
\end{table}

\newpage
\subsection*{Pseudo code of the proposed model}
\begin{algorithm}[H]
  \caption{Simulation Algorithm}\label{alg:simulation}
  \begin{algorithmic}[1]
    \State Define device parameters
    \State Initialise grid (Ngrid)
    \For{each Nparticles}
      \State Distribute particles across the grid
      \State Interpolate charge to grid
    \EndFor
    \For{each Nsteps}
      \State Update Vdevice
      \For{each Ngrid}
        \State $\nabla^2\Phi = -\frac{\rho}{\varepsilon_0}$
        \State $E = -\nabla\Phi$
      \EndFor
      \For{each Nparticles}
        \State Calculate $\nu_D$ (Eq. 1)
        \State Position, $x = x + \Delta t \cdot \nu_D$
      \EndFor
      \State Calculate $d_{\text{eff}}, q(t)$
      \While{error $< 0.0001$}
        \State Apply KVL, KCL (Eq. 2, 3)
        \State Assume $V_{\text{SC,top}} = V_{\text{device}}$
        \State Calculate capacitance parameters (Eqs. 6-8)
        \State Calculate voltage drops, currents: $I_{\text{resistive}} + I_{\text{capacitive}}$ (Eqs. 4, 5, 9)
        \State error $= \frac{I_{\text{SC,top}} - I_{\text{SC,bottom}}}{I_{\text{SC,bottom}}}$
        \State Update $V_{\text{SC,top}}$ based on the error
      \EndWhile
      \If{Nsteps = Nmax}
        \State STOP
      \EndIf
    \EndFor
  \end{algorithmic}
\end{algorithm}

\newpage

\subsection*{Additional plots to support Fig. 4}

\renewcommand{\thefigure}{S1} 
\begin{figure}[!ht]
  \centering
  \begin{subfigure}{0.45\textwidth}
    \includegraphics[width=\linewidth]{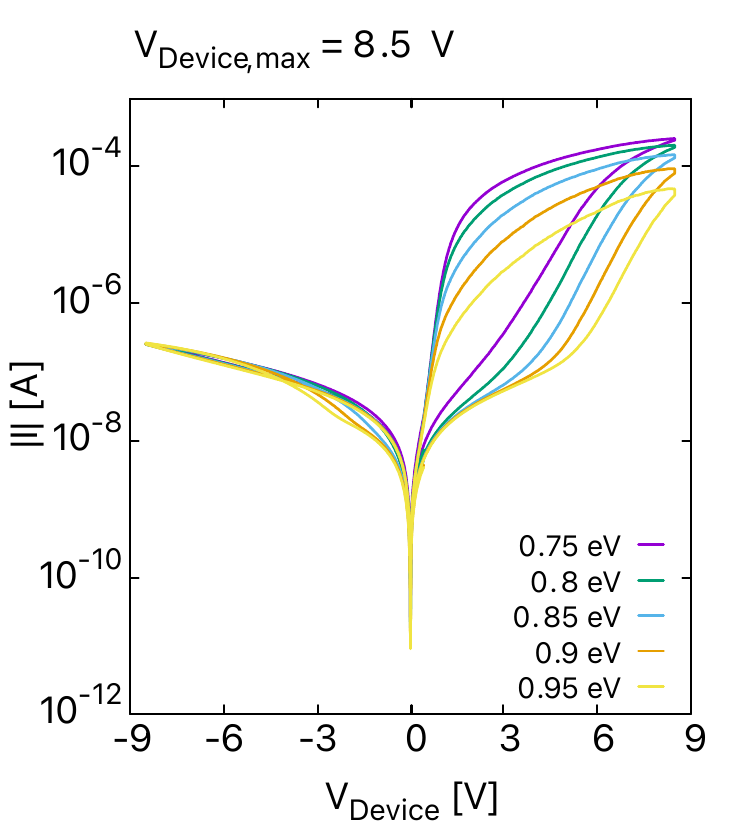}
    \caption{bottom Schottky contact barrier}
  \end{subfigure}
  \hfill
  \begin{subfigure}{0.45\textwidth}
    \includegraphics[width=\linewidth]{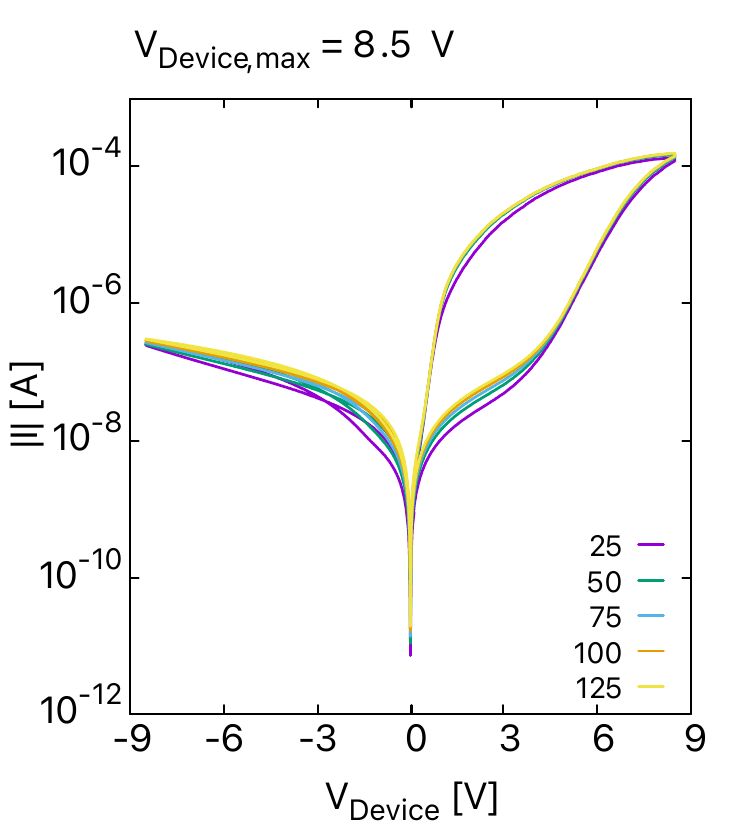}
    \caption{BFO permittivity}
  \end{subfigure}
  \medskip
  \begin{subfigure}{0.45\textwidth}
    \includegraphics[width=\linewidth]{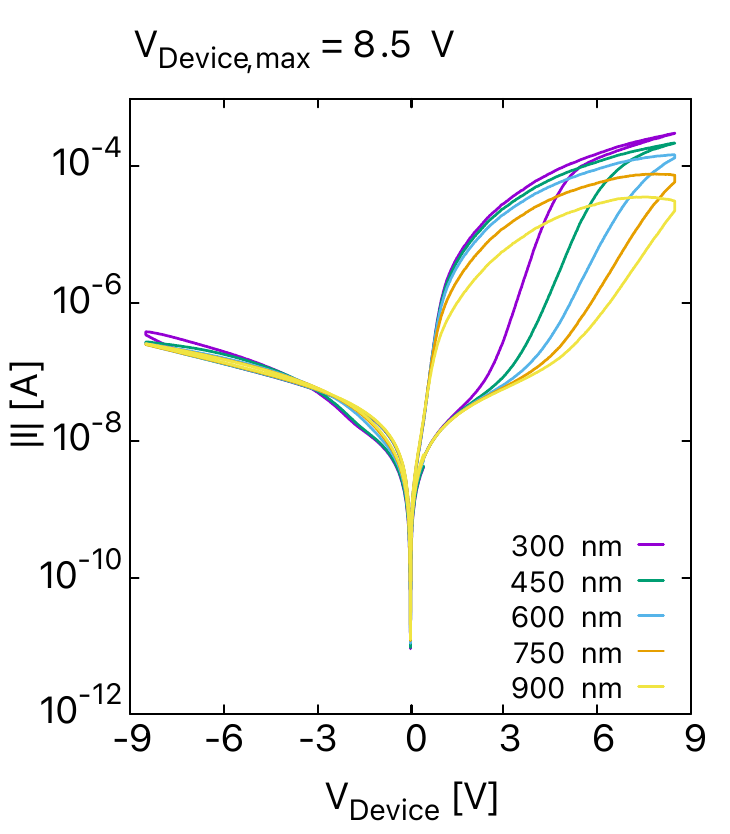}
    \caption{BFO layer length}
  \end{subfigure}
  \hfill
  \begin{subfigure}{0.45\textwidth}
    \includegraphics[width=\linewidth]{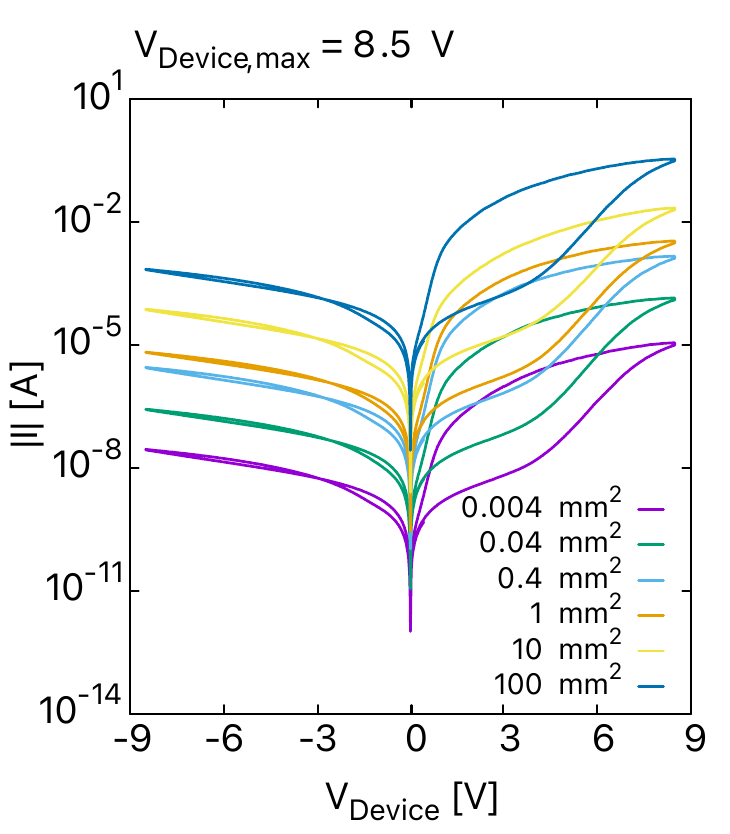}
    \caption{area of the device}
  \end{subfigure}
  \caption{Current-voltage characteristics of BFO device due to change in difference device parameters\cite{Yarragolla2022}}
  \label{fig:S1}
\end{figure}

\newpage 
\renewcommand{\thefigure}{S2} 
\begin{figure}[!ht]
\centering\includegraphics[width=0.7\textwidth]{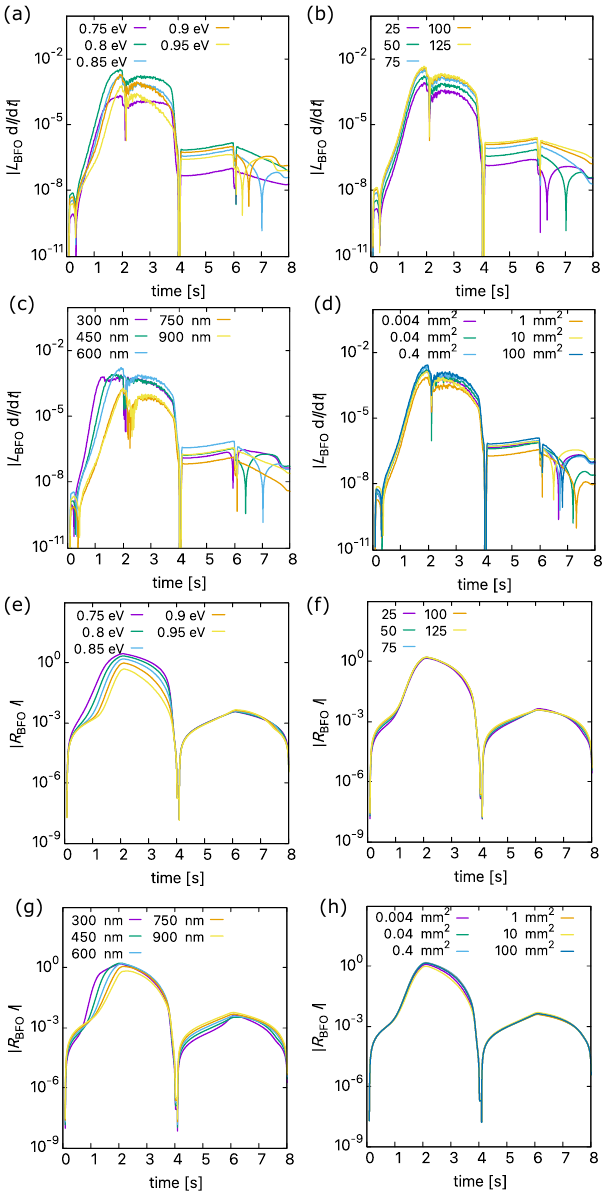}
\caption{For better understanding the inertia effects based on Eq. 13 in the main paper, by using curve fitting tool of Python, the terms (a)-(d) $L_{\rm BFO} \frac{\mathrm{d} I_{\rm BFO}}{\mathrm{d} t}$\,[V] and (e)-(h) $ R_{\rm BFO} I_{\rm BFO}$\,[V] are plotted below. (a),(e): Bottom Schottky contact barrier height; (b),(f): permittivity of oxide layer; (c),(g): length of the oxide layer; (d),(h): area of the device.}
\label{fig:2}
\end{figure}